\documentclass{article}
\usepackage{amsfonts}
\usepackage{makeidx}
\usepackage{latexsym,amsmath,amssymb,amscd}
\usepackage{makecell}
\usepackage{xcolor}
\usepackage{multirow}
\usepackage[all]{xy}
\usepackage{float}
\usepackage{longtable}

\allowdisplaybreaks
\newtheorem{theorem}{Theorem}

\newtheorem{proposition}[theorem]{Proposition}

\begin{document}

\title{Higher order first integrals of autonomous dynamical systems in terms of geometric symmetries}
\author{Antonios Mitsopoulos$^{1,a)}$ and Michael Tsamparlis$^{1,b)}$ \\
%EndAName
{\ \ }\\
$^{1}${\textit{Faculty of Physics, Department of
Astronomy-Astrophysics-Mechanics,}}\\
{\ \textit{University of Athens, Panepistemiopolis, Athens 157 83, Greece}}
\vspace{12pt} %EndAName
\\
$^{a)}$Author to whom correspondence should be addressed: antmits@phys.uoa.gr %EndAName
\\
$^{b)}$Email: mtsampa@phys.uoa.gr }
\date{}
\maketitle

\begin{abstract}
In general, a system of differential equations is integrable if there exist `sufficiently many' first integrals (FIs) so that its solution can be found by means of quadratures. Therefore, the determination of the FIs is an important issue in order to establish the integrability of a dynamical system. In this work, we consider holonomic autonomous dynamical systems defined by equations $\ddot{q}^{a}= -\Gamma_{bc}^{a}(q) \dot{q}^{b}\dot{q}^{c} -Q^{a}(q)$ where $\Gamma^{a}_{bc}(q)$ are the coefficients of a symmetric (possibly non-metrical) connection and $-Q^{a}(q)$ are the generalized forces. We prove a theorem which produces the FIs of any order of such systems in terms of the `symmetries' of the geometry defined by the quantities $\Gamma_{bc}^{a}(q)$. We apply the theorem to compute quadratic and cubic FIs of various dynamical systems.
\end{abstract}

\section{Introduction}

\label{sec.intro}

A first integral (FI) of a second order set of dynamical equations with generalized coordinates $q^{a}$ and generalized velocities $\dot{q}^{a}\equiv \frac{dq^{a}}{dt}$ is a function $I(t,q^{a},\dot{q}^{a})$ satisfying the condition $\frac{dI}{dt}=0$ along the dynamical equations. It is important to have a systematic method for determining FIs because they can be used to reduce the order of the dynamical equations and, if they are `enough' in number \cite{Arnold 1989}, to find the solution of the system by quadrature (Liouville integrability).

The standard method to compute the FIs is the method of Noether symmetries \cite{Damianou 2004, Tsamparlis 2011, Halder 2018}. A different method is the direct method \cite{Katzin 1981, Horwood 2007, TsampMitsA, TsampMitsB, Mits 2021} in which one assumes a functional form for the FI $I$ and demands directly the condition $\frac{dI}{dt}=0$. This condition and the dynamical equations lead to a system of partial differential equations (PDEs) whose solution provides the FIs.

In this work, we apply the direct method to autonomous holonomic dynamical systems in a space with a symmetric connection $\Gamma_{bc}^{a}(q)$ (not necessarily Riemannian) which is read from the dynamical equations. We compute the resulting system of PDEs and solve it in terms of the `symmetries' of $\Gamma_{bc}^{a}(q)$. The result is stated as Theorem \ref{thm1} and provides a systematic method to determine the FIs of any order, time-dependent and autonomous, of these dynamical systems. In the special case where the symmetric connection $\Gamma_{bc}^{a}(q)$ is the Riemannian one defined in terms of the kinetic metric (kinetic energy) $\gamma_{ab}(q)$ of the system, the computed FIs are directly related by means of the Inverse Noether Theorem \cite{TsampMitsB, Djukic} to gauged generalized (i.e. velocity-dependent) Noether symmetries. Finally, we apply Theorem \ref{thm1} in order to find integrable and superintegrable systems that admit quadratic (QFIs) and cubic FIs (CFIs).

\section{The conditions for higher order FIs}

\label{sec.conditions}

We consider autonomous holonomic dynamical systems of the form
\begin{equation}
\ddot{q}^{a}= -\Gamma^{a}_{bc}(q)\dot{q}^{b}\dot{q}^{c} -Q^{a}(q) \label{eq.tk1}
\end{equation}
where $\Gamma^{a}_{bc}(q)$ are the coefficients of a general connection and $-Q^{a}(q)$ are the generalized forces. Since only the symmetric part $\Gamma^{a}_{(bc)}$ contributes to the dynamical equations, without loss of generality, the quantities $\Gamma^{a}_{bc}(q)$ are assumed to be symmetric.

We look for $m$th-order FIs of the general form
\begin{equation}
I^{(m)}= \sum_{r=0}^{m} M_{i_{1}i_{2}...i_{r}}(t,q) \dot{q}^{i_{1}} \dot{q}^{i_{2}}...\dot{q}^{i_{r}}= M+ M_{i_{1}}\dot{q}^{i_{1}} +M_{i_{1}i_{2}} \dot{q}^{i_{1}} \dot{q}^{i_{2}} +...+M_{i_{1}i_{2}...i_{m}}\dot{q}^{i_{1}} \dot{q}^{i_{2}} ...\dot{q}^{i_{m}} \label{FI.5}
\end{equation}
where $M_{i_{1}...i_{r}}(t,q)$ are totally symmetric $r$-rank tensors and the index $(m)$ denotes the order of the FI.

The FI condition
\begin{equation}
\frac{dI}{dt}=0  \label{DS1.10a}
\end{equation}%
and the dynamical equations (\ref{eq.tk1}) result in the following system of PDEs:
\begin{eqnarray}
M_{(i_{1}i_{2}...i_{m}|i_{m+1})} &=&0  \label{eq.veldep4.1} \\
M_{i_{1}i_{2}...i_{m},t}+M_{(i_{1}i_{2}...i_{m-1}|i_{m})} &=&0
\label{eq.veldep4.2} \\
M_{i_{1}i_{2}...i_{r},t}+M_{(i_{1}i_{2}...i_{r-1}|i_{r})} -(r+1)M_{i_{1}i_{2}...i_{r}i_{r+1}}Q^{i_{r+1}} &=&0, \enskip r=1,2,...,m-1 \label{eq.veldep4.3} \\
M_{,t}-M_{i_{1}}Q^{i_{1}} &=&0 \label{eq.veldep4.4} \\
M_{i_{1},tt}-2M_{i_{1}i_{2},t}Q^{i_{2}}+\left( M_{c}Q^{c}\right) _{,i_{1}} &=&0  \label{eq.int1} \\
2\left( M_{[i_{1}\wr c \wr}Q^{c}\right) _{|i_{2}]}-M_{[i_{1}|i_{2}],t} &=&0 \label{eq.int2}
\end{eqnarray}
where $|$ denotes the covariant derivative with respect to (wrt) the symmetric connection $\Gamma^{a}_{bc}$, a comma indicates partial derivative wrt $q^{a}$ or $t$, round/square brackets indicate symmetrization/antisymmetrization of the enclosed indices, and indices enclosed between wavy lines are overlooked by symmetrization or antisymmetrization symbols.

Equations (\ref{eq.int1}) and (\ref{eq.int2}) express the integrability conditions $M_{,i_{1}t}=M_{,ti_{1}}$ and $M_{,[i_{1}i_{2}]}=0$ of the scalar $M$, respectively.

Equation (\ref{eq.veldep4.1}) generalizes the concept of Killing tensors (KTs) to a non-metrical geometry with a symmetric connection $\Gamma^{a}_{bc}$. In this context, $M_{i_{1}i_{2}...i_{m}}$ is a \textbf{generalized $\mathbf{m}$th-order KT} for $\Gamma^{a}_{bc}$.

The most general choice for the generalized $m$th-order KT $M_{i_{1}i_{2}...i_{m}}$ in the case of an autonomous system is
\begin{equation}
M_{i_{1}...i_{m}}(t,q)=C_{(0)i_{1}...i_{m}}(q)+%
\sum_{N=1}^{n}C_{(N)i_{1}...i_{m}}(q)\frac{t^{N}}{N}
\label{eq.aspm1}
\end{equation}%
where $C_{(N)i_{1}...i_{m}}(q)$, $N=0,1,...,n$, is a sequence of arbitrary $m$th-order generalized KTs of $\Gamma^{a}_{bc}$ and $n$ is the degree of the considered polynomial.

The choice (\ref{eq.aspm1}) and equation (\ref{eq.veldep4.2}) indicate that we set
\begin{equation}
M_{i_{1}...i_{r}}(t,q)=\sum_{N_{r}=0}^{n_{r}} L_{(N_{r})i_{1}...i_{r}}(q)t^{N_{r}},\enskip r=1,2,...,m-1  \label{eq.aspm2}
\end{equation}%
where $L_{(N_{r})i_{1}...i_{r}}(q)$, $N_{r}=0,1,...,n_{r}$, are arbitrary $r$-rank totally symmetric tensors and $n_{r}$ is the
degree of the considered polynomial.

We note that the degrees $n, n_{r}$ of the above polynomial expressions of $t$ may be infinite.

Substituting (\ref{eq.aspm1}) and (\ref{eq.aspm2}) in the system of equations (\ref{eq.veldep4.1}) - (\ref{eq.int2}) (eq. (\ref{eq.veldep4.1}) is identically zero since $C_{(N)i_{1}...i_{m}}$ are assumed to be generalized KTs), we end up with a system of five PDEs. The solution of this system is lengthy and requires the consideration of many cases and subcases. We state the solution below as Theorem \ref{thm1}.

\section{Theorem for $m$th-order FIs of an autonomous holonomic dynamical system}

\label{sec.theorem}

\begin{theorem} \label{thm1}
The independent $m$th-order FIs of the dynamical system (\ref{eq.tk1}) are the following:
\bigskip

\textbf{Integral 1.}
\begin{eqnarray*}
I_{n}^{(m)} &=&\left( -\sum^{n}_{N=1} \frac{t^{N}}{N}L_{(N-1)(i_{1}...i_{m-1}|i_{m})} +C_{(0)i_{1}...i_{m}}\right) \dot{q}^{i_{1}} ... \dot{q}^{i_{m}}+ \sum_{r=1}^{m-1} \left( \sum^{n}_{N=0}
t^{N}L_{(N)i_{1}...i_{r}} \right) \dot{q}^{i_{1}} ... \dot{q}^{i_{r}} + \notag \\
&& +s\frac{t^{n+1}}{n+1} +\sum^{n}_{N=1} L_{(N-1)c}Q^{c}\frac{t^{N}}{N} +G(q)
\end{eqnarray*}%
where $C_{(0)i_{1}...i_{m}}$, $L_{(N)(i_{1}...i_{m-1}|i_{m})}$ for $%
N=0,1,...,n-1$ are $\mathbf{m}${\textbf{th-order generalized KTs}}, $L_{(n)i_{1}...i_{m-1}}$ is an $\mathbf{(m-1)}${\textbf{th-order generalized KT}}, $s$ is an arbitrary constant defined by the
condition
\begin{equation}
L_{(n)i_{1}}Q^{i_{1}}=s  \label{eq.FI1f}
\end{equation}%
while the vectors $L_{(N)i_{1}}$ and the {\textbf{totally symmetric tensors}} $%
L_{(A)i_{1}...i_{r}}$, $A=0,1,...,n$, $r=2,3,...,m-2$ satisfy the conditions:
\begin{eqnarray}
L_{(n)(i_{1}...i_{m-2}|i_{m-1})} &=&-\frac{m}{n}%
L_{(n-1)(i_{1}...i_{m-1}|i_{m})}Q^{i_{m}}  \label{eq.FI1a} \\
L_{(k-1)(i_{1}...i_{m-2}|i_{m-1})} &=&-\frac{m}{k-1}%
L_{(k-2)(i_{1}...i_{m-1}|i_{m})}Q^{i_{m}}-kL_{(k)i_{1}...i_{m-1}}, \enskip k=2,3,...,n  \label{eq.FI1b} \\
L_{(0)(i_{1}...i_{m-2}|i_{m-1})}
&=&mC_{(0)i_{1}...i_{m-1}i_{m}}Q^{i_{m}}-L_{(1)i_{1}...i_{m-1}}
\label{eq.FI1c} \\
L_{(n)(i_{1}...i_{r-1}|i_{r})}
&=&(r+1)L_{(n)i_{1}...i_{r}i_{r+1}}Q^{i_{r+1}}, \enskip r=2,3,...,m-2 \label{eq.FI1d} \\
L_{(k-1)(i_{1}...i_{r-1}|i_{r})}
&=&(r+1)L_{(k-1)i_{1}...i_{r}i_{r+1}}Q^{i_{r+1}} -kL_{(k)i_{1}...i_{r}}, \enskip k=1,2,...,n, \enskip r=2,3,...,m-2 \label{eq.FI1e} \\
\left( L_{(n-1)c}Q^{c}\right) _{,i_{1}} &=&2nL_{(n)i_{1}i_{2}}Q^{i_{2}}
\label{eq.FI1g} \\
\left( L_{(k-2)c}Q^{c}\right) _{,i_{1}}
&=&2(k-1)L_{(k-1)i_{1}i_{2}}Q^{i_{2}}-k(k-1)L_{(k)i_{1}}, \enskip k=2,3,...,n \label{eq.FI1h} \\
G_{,i_{1}} &=&2L_{(0)i_{1}i_{2}}Q^{i_{2}}-L_{(1)i_{1}}.  \label{eq.FI1i}
\end{eqnarray}

\textbf{Integral 2.}
\begin{equation*}
I^{(m)}_{e}= \frac{e^{\lambda t}}{\lambda} \left(
-L_{(i_{1}...i_{m-1}|i_{m})} \dot{q}^{i_{1}} ... \dot{q}^{i_{m}} + \lambda
\sum_{r=1}^{m-1} L_{i_{1}...i_{r}} \dot{q}^{i_{1}} ... \dot{q}^{i_{r}} +
L_{i_{1}}Q^{i_{1}} \right)
\end{equation*}
where $\lambda\neq0$, $L_{(i_{1}...i_{m-1}|i_{m})}$ is an $m$th-order generalized KT and the remaining totally symmetric tensors satisfy the conditions:
\begin{eqnarray}
L_{(i_{1}...i_{m-2}|i_{m-1})}&=& -\frac{m}{\lambda}
L_{(i_{1}...i_{m-1}|i_{m})} Q^{i_{m}} -\lambda L_{i_{1}...i_{m-1}}
\label{eq.FI2a} \\
L_{(i_{1}...i_{r-1}|i_{r})}&=&(r+1) L_{i_{1}...i_{r}i_{r+1}} Q^{i_{r+1}}
-\lambda L_{i_{1}...i_{r}}, \enskip r=2,3,...,m-2  \label{eq.FI2b} \\
\left(L_{c}Q^{c}\right)_{,i_{1}}&=& 2\lambda L_{i_{1}i_{2}} Q^{i_{2}}
-\lambda^{2}L_{i_{1}}.  \label{eq.FI2c}
\end{eqnarray}
\end{theorem}

We note that Theorem \ref{thm1} for $m=2$ and a Riemannian connection reduces to Theorem 1 of \cite{TsampMitsA} and to Theorem 3 of \cite{TsampMitsB} for the case of QFIs.

Moreover, we have the following minor results.

\begin{proposition} \label{pro1}
The independent $m$th-order FIs $I_{n}^{(m)}$ and $I^{(m)}_{e}$ satisfy the following recursion formulae:\newline
a. $I_{n}^{(k)}<I_{n}^{(k+1)}$, that is, each $k$th-order FI $I_{n}^{(k)}$ is a subcase of the next $(k+1)$th-order FI $I_{n}^{(k+1)}$ with the same degree $n$ of time-dependence for all $k\in \mathbb{N}$. \newline
b. $I_{\ell}^{(m)}<I_{\ell+1}^{(m)}$, that is, the $m$th-order FI $I_{\ell}^{(m)}$ with time-dependence fixed by $\ell$ is a subcase of the $m$th-order FI $I_{\ell+1}^{(m)}$ with time-dependence $\ell+1$ for all $\ell\in \mathbb{N}$. \newline
c. $I_{e}^{(k)}<I_{e}^{(k+1)}$, that is, each $k$th-order FI $I_{e}^{(k)}$ is a subcase of the next $(k+1)$th-order FI $I_{e}^{(k+1)}$ for all $k\in \mathbb{N}$.
\end{proposition}

\begin{proposition} \label{pro2}
The $m$th-order FI $I^{(m)}_{n}$ consists of the following two independent FIs:\newline
a. The FI $J^{(m,1)}_{\ell}$ whose coefficients are polynomials of $t$ containing even powers of $t$ for even products of velocities and odd powers of $t$ for odd products of velocities. \newline
b. The FI $J^{(m,2)}_{\ell}$ whose coefficients are polynomials of $t$ containing even powers of $t$ for odd products of velocities and odd powers of $t$ for even products of velocities.
\end{proposition}

For even orders $m=2\nu$ ($\nu\in\mathbb{N}$) the independent FIs of Proposition \ref{pro2} are computed by the formulae ($\ell\in\mathbb{N}$):

a.
\begin{eqnarray}
J^{(m=2\nu,1)}_{\ell}&=& \left( -\frac{t^{2\ell}}{2\ell} L_{(2\ell-1)(i_{1}...i_{m-1}|i_{m})} - ... - \frac{t^{2}}{2} L_{(1)(i_{1}...i_{m-1}|i_{m})} +C_{(0)i_{1}...i_{m}} \right) \dot{q}^{i_{1}} ... \dot{q}^{i_{m}} + \notag \\
&& + \sum_{1\leq r \leq m-1}^{odd} \left( t^{2\ell-1} L_{(2\ell-1)i_{1}...i_{r}} + ... + t^{3}L_{(3)i_{1}...i_{r}} +tL_{(1)i_{1}...i_{r}} \right) \dot{q}^{i_{1}} ... \dot{q}^{i_{r}} + \notag \\
&& + \sum_{1\leq r \leq m-1}^{even} \left( t^{2\ell} L_{(2\ell)i_{1}...i_{r}} + ... + t^{2}L_{(2)i_{1}...i_{r}} +L_{(0)i_{1}...i_{r}} \right) \dot{q}^{i_{1}} ... \dot{q}^{i_{r}} + \notag \\
&& + \frac{t^{2\ell}}{2\ell}L_{(2\ell-1)c}Q^{c} + ... + \frac{t^{2}}{2}L_{(1)c}Q^{c} + G(q) \label{eq.FI4a}
\end{eqnarray}
where $C_{(0)i_{1}...i_{m}}$, $L_{(N)(i_{1}...i_{m-1}|i_{m})}$ for $N=1,3,...,2\ell-1$ are $m$th-order generalized KTs and the following conditions are satisfied:
\begin{eqnarray}
L_{(2\ell)(i_{1}...i_{m-2}|i_{m-1})}&=& -\frac{m}{2\ell} L_{(2\ell-1)(i_{1}...i_{m-1}|i_{m})} Q^{i_{m}} \label{eq.FI4.1} \\
L_{(k-1)(i_{1}...i_{m-2}|i_{m-1})}&=&-\frac{m}{k-1} L_{(k-2)(i_{1}...i_{m-1}|i_{m})} Q^{i_{m}} -kL_{(k)i_{1}...i_{m-1}}, \enskip k=3,5,...,2\ell-1 \label{eq.FI4.2} \\
L_{(0)(i_{1}...i_{m-2}|i_{m-1})}&=& mC_{(0)i_{1}...i_{m-1}i_{m}} Q^{i_{m}} -L_{(1)i_{1}...i_{m-1}} \label{eq.FI4.3} \\
L_{(2\ell)(i_{1}...i_{r-1}|i_{r})}&=& (r+1) L_{(2\ell)i_{1}...i_{r}i_{r+1}} Q^{i_{r+1}}, \enskip r=3,5,...,m-3 \label{eq.FI4.4} \\ L_{(k-1)(i_{1}...i_{r-1}|i_{r})}&=& (r+1) L_{(k-1)i_{1}...i_{r}i_{r+1}} Q^{i_{r+1}} -kL_{(k)i_{1}...i_{r}}, \enskip k=1,3,...,2\ell-1, r=3,5,...,m-3 \notag \\
\label{eq.FI4.5} \\
L_{(k-1)(i_{1}...i_{r-1}|i_{r})}&=& (r+1) L_{(k-1)i_{1}...i_{r}i_{r+1}} Q^{i_{r+1}} -kL_{(k)i_{1}...i_{r}}, \enskip k=2,4,...,2\ell, \enskip r=2,4,...,m-2 \label{eq.FI4.6} \\
\left( L_{(2\ell-1)c}Q^{c} \right)_{,i_{1}} &=& 4\ell L_{(2\ell)i_{1}i_{2}}Q^{i_{2}} \label{eq.FI4.7} \\
\left( L_{(k-2)c}Q^{c} \right)_{,i_{1}} &=& 2(k-1)L_{(k-1)i_{1}i_{2}}Q^{i_{2}} -k(k-1)L_{(k)i_{1}}, \enskip k=3,5,...,2\ell-1 \label{eq.FI4.8} \\
G_{,i_{1}}&=& 2L_{(0)i_{1}i_{2}}Q^{i_{2}} -L_{(1)i_{1}}. \label{eq.FI4.9}
\end{eqnarray}

b.
\begin{eqnarray}
J^{(m=2\nu,2)}_{\ell}&=& \left( -\frac{t^{2\ell+1}}{2\ell+1} L_{(2\ell)(i_{1}...i_{m-1}|i_{m})} - ... - \frac{t^{3}}{3} L_{(2)(i_{1}...i_{m-1}|i_{m})} - tL_{(0)(i_{1}...i_{m-1}|i_{m})} \right) \dot{q}^{i_{1}} ... \dot{q}^{i_{m}} + \notag \\
&& + \sum_{1\leq r \leq m-1}^{odd} \left( t^{2\ell} L_{(2\ell)i_{1}...i_{r}} + ... + t^{2}L_{(2)i_{1}...i_{r}} +L_{(0)i_{1}...i_{r}} \right) \dot{q}^{i_{1}} ... \dot{q}^{i_{r}}+ \notag \\
&& + \sum_{1\leq r \leq m-1}^{even} \left( t^{2\ell+1} L_{(2\ell+1)i_{1}...i_{r}} + ... + t^{3}L_{(3)i_{1}...i_{r}} +tL_{(1)i_{1}...i_{r}} \right) \dot{q}^{i_{1}} ... \dot{q}^{i_{r}} + \notag \\
&& + \frac{t^{2\ell+1}}{2\ell+1}L_{(2\ell)c}Q^{c} + ... + \frac{t^{3}}{3}L_{(2)c}Q^{c} + tL_{(0)c}Q^{c} \label{eq.FI4b}
\end{eqnarray}
where $L_{(N)(i_{1}...i_{m-1}|i_{m})}$ for $N=0,2,...,2\ell$ are $m$th-order generalized KTs and the following conditions are satisfied:
\begin{eqnarray}
L_{(2\ell+1)(i_{1}...i_{m-2}|i_{m-1})}&=& -\frac{m}{2\ell+1} L_{(2\ell)(i_{1}...i_{m-1}|i_{m})} Q^{i_{m}} \label{eq.FI5.1} \\ L_{(k-1)(i_{1}...i_{m-2}|i_{m-1})}&=& -\frac{m}{k-1} L_{(k-2)(i_{1}...i_{m-1}|i_{m})} Q^{i_{m}} -kL_{(k)i_{1}...i_{m-1}}, \enskip k=2,4,...,2\ell \label{eq.FI5.2} \\
L_{(2\ell+1)(i_{1}...i_{r-1}|i_{r})}&=& (r+1) L_{(2\ell+1)i_{1}...i_{r}i_{r+1}} Q^{i_{r+1}}, \enskip r=3,5,...,m-3 \label{eq.FI5.3} \\
L_{(k-1)(i_{1}...i_{r-1}|i_{r})}&=& (r+1) L_{(k-1)i_{1}...i_{r}i_{r+1}} Q^{i_{r+1}} -kL_{(k)i_{1}...i_{r}}, \enskip k=1,3,...,2\ell+1, r=2,4,...,m-2 \notag \\
\label{eq.FI5.4} \\
L_{(k-1)(i_{1}...i_{r-1}|i_{r})}&=& (r+1) L_{(k-1)i_{1}...i_{r}i_{r+1}} Q^{i_{r+1}} -kL_{(k)i_{1}...i_{r}}, \enskip k=2,4,...,2\ell, \enskip r=3,5...,m-3 \label{eq.FI5.5} \\
\left( L_{(2\ell)c}Q^{c} \right)_{,i_{1}} &=& 2(2\ell+1)L_{(2\ell+1)i_{1}i_{2}}Q^{i_{2}} \label{eq.FI5.6} \\
\left( L_{(k-2)c}Q^{c} \right)_{,i_{1}} &=& 2(k-1)L_{(k-1)i_{1}i_{2}}Q^{i_{2}} -k(k-1)L_{(k)i_{1}}, \enskip k=2,4,...,2\ell. \label{eq.FI5.7}
\end{eqnarray}

Moreover, for even order FIs, it holds that
\begin{eqnarray*}
I^{(2\nu)}_{2k}&=& J^{(2\nu,1)}_{k} + J^{(2\nu,2)}_{k} \left( L_{(2k)(i_{1}...i_{m-1}|i_{m})}=0; L_{(2k+1)(i_{1}...i_{r})}=0, 1\leq r \leq m-1, r=even \right) \\
I^{(2\nu)}_{2k+1}&=& J^{(2\nu,1)}_{k+1}\left( L_{(2k+1)(i_{1}...i_{m-1}|i_{m})}=0; L_{(2k+2)(i_{1}...i_{r})}=0, 1\leq r \leq m-1, r=even \right) + J^{(2\nu,2)}_{k}
\end{eqnarray*}
where $m=2\nu$, while for odd order FIs
\begin{eqnarray*}
I^{(2\nu+1)}_{2k}&=& J^{(2\nu+2,1)}_{k} \left(M_{i_{1}...i_{m}}=0\right) + J^{(2\nu+2,2)}_{k} \left( M_{i_{1}...i_{m}}=0; L_{(2k+1)(i_{1}...i_{r})}=0, 1\leq r \leq m-1, r=even \right) \\
I^{(2\nu+1)}_{2k+1}&=& J^{(2\nu+2,1)}_{k+1}\left( M_{i_{1}...i_{m}}=0; L_{(2k+2)(i_{1}...i_{r})}=0, 1\leq r \leq m-1, r=even \right) + J^{(2\nu+2,2)}_{k}\left(M_{i_{1}...i_{m}}=0\right)
\end{eqnarray*}
where $m=2\nu+2$. For completeness, we may introduce the notation $J^{(2\nu+1,1)}_{\ell} \equiv J^{(2\nu+2,1)}_{\ell}\left(M_{i_{1}...i_{m}}=0\right)$ and $J^{(2\nu+1,2)}_{\ell} \equiv J^{(2\nu+2,2)}_{\ell}\left(M_{i_{1}...i_{m}}=0\right)$.

In the case of a Riemannian connection, the general $m$th-order FIs (\ref{FI.5}) are related to the generalized gauged weak Noether symmetry
\begin{equation}
\left( \xi=0, \enskip \eta_{i_{1}}= -\frac{\partial I^{(m)}}{\partial \dot{q}^{i_{1}}}, \enskip \phi_{a}, \enskip f =I^{(m)} -\frac{\partial I^{(m)}}{\partial \dot{q}^{i_{1}}}\dot{q}^{i_{1}} \right) \quad \text{such that $\phi_{a} \dot{q}^{a} +F^{a} \frac{\partial I^{(m)}}{\partial \dot{q}^{a}}= 0$} \label{eq.weak6}
\end{equation}
by means of the Inverse Noether Theorem \cite{TsampMitsB, Djukic}. In (\ref{eq.weak6}), we have
\[
\frac{\partial I^{(m)}}{\partial \dot{q}^{i_{1}}} = M_{i_{1}} + 2M_{i_{1}i_{2}}\dot{q}^{i_{2}} + 3M_{i_{1}i_{2}i_{3}} \dot{q}^{i_{2}} \dot{q}^{i_{3}} + ... + m M_{i_{1}i_{2}...i_{m}} \dot{q}^{i_{2}}...\dot{q}^{i_{m}} = \sum^{m-1}_{r=0} (r+1) M_{i_{1}i_{2}...i_{r+1}} \dot{q}^{i_{2}}...\dot{q}^{i_{r+1}}
\]
$F^{a}(t,q,\dot{q})$ are the non-conservative generalized forces, $\phi^{a}(t,q,\dot{q})$ is an additional vector generator, $f(t,q,\dot{q})$ is the Noether function, and $\mathbf{X}= \xi(t,q,\dot{q}) \partial_{t} + \eta^{a}(t,q,\dot{q}) \partial_{q^{a}}$ is the Lie generator. In the following, we consider applications which show the importance of Theorem 1.

\section{Applications}

\subsection{Application 1: The QFIs of a non-Riemannian dynamical system}

Consider the dynamical system:
\begin{eqnarray}
\ddot{u} &=& -\frac{8\beta }{u^{3}}\left( u\dot{u}\dot{w} -w\dot{u}^{2}\right) -\frac{1}{u^{2}} \label{eq.aham8a} \\
\ddot{w} &=& -\frac{4\beta }{u^{3}}\left( u\dot{w}^{2} -4w\dot{u}\dot{w}\right) +\frac{2w}{u^{3}} \label{eq.aham8b}
\end{eqnarray}
where $\beta$ is an arbitrary real constant. This system is autonomous holonomic of the form (\ref{eq.tk1}) with variables
\[
q^{a}=
\left(
  \begin{array}{c}
    u \\
    w \\
  \end{array}
\right), \enskip
Q^{a}= \frac{1}{u^{2}}
\left(
  \begin{array}{c}
    1 \\
    -\frac{2w}{u} \\
  \end{array}
\right).
\]
The symmetric connection coefficients are read from the the dynamical equations:
\begin{equation}
\Gamma^{1}_{22}=\Gamma^{2}_{11}=0, \enskip \Gamma^{1}_{11}= \Gamma^{2}_{12}= -8\beta\frac{w}{u^{3}}, \enskip \Gamma^{1}_{12}= \Gamma^{2}_{22}= \frac{4\beta}{u^{2}}. \label{eq.conne}
\end{equation}

The curvature tensor $R^{a}{}_{bcd}= \Gamma^{a}_{bd,c} -\Gamma^{a}_{bc,d} +\Gamma^{a}_{sc}\Gamma^{s}_{bd} -\Gamma^{a}_{sd}\Gamma^{s}_{bc}$ is computed to be
\[
R^{1}{}_{112}= R^{2}{}_{221} = -R^{2}{}_{212} = -R^{1}{}_{121}= -\frac{32b^{2}w}{u^{5}}, \enskip R^{2}{}_{112}= -R^{2}{}_{121}= \frac{24bw}{u^{4}}.
\]

Solving the generalized KT condition $C_{(ab|c)}=0$, we find that the connection (\ref{eq.conne}) admits only the second order generalized KT
\begin{equation}
C_{ab}= k e^{\frac{12\beta w}{u^{2}}}
\left(
  \begin{array}{cc}
    0 & 1 \\
    1 & 0 \\
  \end{array}
\right) \label{eq.KT1}
\end{equation}
where $k$ is an arbitrary constant.

Solving the generalized Killing vector (KV) condition $L_{(a|b)}=0$, we find $L_{a}=0$; therefore, generalized KVs do not exist.

Moreover, it can be shown that non-zero vectors $B_{a}$ which generate reducible generalized KTs of the form $B_{(a|b)}$ do not exist as well.

Applying Theorem \ref{thm1} for $m=2$, we find that the system admits only one QFI which is the
\begin{equation}
J^{(2,1)}_{1}= e^{\frac{12\beta w}{u^{2}}} \left( \dot{u}\dot{w} +\frac{1}{12\beta} \right). \label{eq.qint1.5}
\end{equation}
To prove that the given system is integrable, we need one more independent FI of higher order in involution.

\subsection{Application 2: CFIs of a class of autonomous conservative dynamical systems.}

Consider the two-dimensional (2d) autonomous conservative dynamical systems with potential $V=F(x^{2}+\nu y^{2})$ where $\nu$ is an arbitrary constant. Determine the potentials which admit CFIs.

In \cite{Fokas}, the authors establish an isomorphism between the autonomous QFIs/CFIs of Hamilton's equations of an autonomous conservative dynamical system and the admissible Lie-B\"{a}cklund symmetries of the Hamilton-Jacobi equation. Using this result, they found the following three such potentials:
\begin{equation}
V_{(1a)}=\frac{1}{2}x^{2}+\frac{9}{2}y^{2},\enskip V_{(1b)}=\frac{1}{2}x^{2}+\frac{1}{18}y^{2},\enskip V_{(1c)}=(x^{2}-y^{2})^{-2/3}.  \label{Fok1}
\end{equation}%

Applying Theorem \ref{thm1}, we find two more potentials and  show that the results of \cite{Fokas} are just special cases. These potentials are the following:
\bigskip

a. The new superintegrable potential
\begin{equation}
V_{1}= c_{0}(x^{2} +9y^{2}) +c_{1}y \label{Fok2}
\end{equation}
where $c_{0}$ and $c_{1}$ are arbitrary constants, which admits the associated CFI
\begin{equation}
J_{1}= (x\dot{y} -y\dot{x})\dot{x}^{2} -\frac{c_{1}}{18c_{0}} \dot{x}^{3} +\frac{c_{1}}{3}x^{2}\dot{x} +6c_{0}x^{2}y\dot{x} - \frac{2c_{0}}{3}x^{3}\dot{y}. \label{Fok3}
\end{equation}

b. The integrable potential
\begin{equation}
V_{2}= k(x^{2}-y^{2})^{-2/3} \label{Fok4}
\end{equation}
where $k$ is an arbitrary constant, which admits the CFI
\begin{equation}
J_{2}= \left( x\dot{y} -y\dot{x} \right) \left( \dot{y}^{2} -\dot{x}^{2} \right) +4V_{2}(y\dot{x} +x\dot{y}). \label{Fok6}
\end{equation}

The potentials (\ref{Fok1}) are special cases of $V_{1}$ and $V_{2}$ for the following values of the parameters:
\[
V_{(1a)}= V_{1}\left( c_{1}=0, c_{0}=\frac{1}{2} \right),
\enskip V_{(1b)}= V_{1}\left( x \leftrightarrow y; c_{1}=0, c_{0}=\frac{1}{18} \right), \enskip V_{(1c)} =V_{2}(k=1).
\]

\subsection{Application 3: New integrable/superintegrable potentials that admit autonomous and time-dependent CFIs.}

In \cite{Karlovini 2000}, the authors using the Jacobi metric approach found integrable and superintegrable potentials that admit autonomous CFIs.

Applying Theorem \ref{thm1}, we find the new integrable potential
\begin{equation}
V= \frac{k_{1}}{(a_{2}y-a_{5}x)^{2}} +\frac{k_{2}}{r} + \frac{k_{3}(a_{2}x+a_{5}y)}{r(a_{2}y-a_{5}x)^{2}} \label{new1}
\end{equation}
where $k_{1}, k_{2}, k_{3}, a_{2}, a_{5}$ are arbitrary constants and $r=\sqrt{x^{2}+y^{2}}$, which admits the CFI
\begin{eqnarray*}
J_{1} &=&(x\dot{y}-y\dot{x})^{2}(a_{2}\dot{x}+a_{5}\dot{y}) +\frac{2k_{1}r^{2}%
}{(a_{2}y-a_{5}x)^{2}}(a_{2}\dot{x}+a_{5}\dot{y}) -\frac{k_{2}(a_{2}y-a_{5}x)%
}{r}(x\dot{y}-y\dot{x}) + \\
&& +\frac{k_{3}r}{a_{2}y-a_{5}x}(a_{2}\dot{y} -a_{5}\dot{x}) -\frac{k_{3}(a_{2}x+a_{5}y)}{r(a_{2}y-a_{5}x)}(x\dot{y}-y\dot{x}) +\frac{2k_{3}(a_{2}x+a_{5}y)r}{(a_{2}y-a_{5}x)^{2}}(a_{2}\dot{x} +a_{5}\dot{y}).
\end{eqnarray*}

We note that for $k_{2}=0$, the special potential
\begin{equation}
V(k_{2}=0)= \frac{k_{1}}{(a_{2}y-a_{5}x)^{2}} + \frac{k_{3}(a_{2}x+a_{5}y)}{r(a_{2}y-a_{5}x)^{2}} \label{eq.su1}
\end{equation}
admits also the additional time-dependent CFI
\begin{eqnarray*}
J_{2}&=& -tJ_{1}(k_{2}=0) +(a_{2}x +a_{5}y)(x\dot{y} -y\dot{x})^{2} + \frac{2k_{1}r^{2}(a_{2}x +a_{5}y)}{(a_{2}y -a_{5}x)^{2}} + \\ &&+\frac{2k_{3}r(a_{2}x +a_{5}y)^{2}}{(a_{2}y -a_{5}x)^{2}} +k_{3}r.
\end{eqnarray*}
We conclude that (\ref{eq.su1}) is a new superintegrable potential. This result illustrates the importance of the time-dependent FIs in the determination of the integrability/superintegrability.

\section{Conclusions}

\label{sec.conclusions}

We draw the following conclusions: \newline
a) We have developed a direct systematic method to compute the $m$th-order FIs of the autonomous holonomic dynamical systems (\ref{eq.tk1}) in terms of the `symmetries' of the geometric objects (symmetric connection or kinetic metric, depending on the case) defined by the dynamical equations. \newline
b) This method applies to non-Riemanian geometries with a symmetric connection. It has been shown that the $m$th-order FIs require the generalized KTs and KVs defined by the symmetric connection $\Gamma _{bc}^{a}$. The case of a Riemannian connection is just a special case. \newline
c) The system of PDEs (\ref{eq.veldep4.1}) - (\ref{eq.int2}) resulting from the condition $dI/dt=0$ and the dynamical equations consists of two parts: A geometric part (eqs. (\ref{eq.veldep4.1}), (\ref{eq.veldep4.2}) ) common to all systems which share the same connection; and a dynamical part (eqs. (\ref{eq.veldep4.3}) - (\ref{eq.int2}) ) which includes the generalized forces $Q^{a}$ of the specific system. \newline
d) All $m$th-order FIs of an autonomous holonomic dynamical system in a
Riemannian background geometry via the Inverse Noether Theorem are
associated to a gauged weak Noether symmetry.

\end{document}